\documentclass[prd,onecolumn,notitlepage,showpacs,preprintnumbers,amsmath,amssymb,nofootinbib,APS,10pt,superscriptaddress]{revtex4-1}

\usepackage{dcolumn}
\usepackage{bm}
\usepackage{xcolor}
\usepackage[utf8]{inputenc}
\usepackage[spanish,english]{babel}
\usepackage{amsmath,amssymb,amsfonts,latexsym,cancel}
\usepackage{graphicx}
\usepackage{color}
\usepackage{soul}
\usepackage{ulem}
\usepackage{hyperref}
\usepackage{amsmath}
\usepackage{slashed}
\usepackage{braket}
\usepackage{graphicx,amssymb,amsmath,amsthm,amsfonts,epsfig,epsf}

\allowdisplaybreaks

\newcommand{\be}{\begin{equation}}
\newcommand{\ee}{\end{equation}}
\newcommand{\bea}{\begin{eqnarray}}
\newcommand{\eea}{\end{eqnarray}}

\newcommand{\black}{\textcolor{black}}

\begin{document}
\title{Quantum corrections to the Schwarzschild metric from vacuum polarization}

\author{Pau Beltr\'an-Palau}\email{pau.beltran@uv.es}
\affiliation{Departamento de Fisica Teorica and IFIC, Centro Mixto Universidad de Valencia-CSIC. Facultad de Fisica, Universidad de Valencia, Burjassot-46100, Valencia, Spain.}
\author{Adrián del Río}\email{axd570@psu.edu}
\affiliation{Institute for Gravitation and the Cosmos \& Physics Department,The Pennsylvania State University, University Park, Pennsylvania 16802 USA}
\author{Jos\'e Navarro-Salas}\email{jnavarro@ific.uv.es}
\affiliation{Departamento de Fisica Teorica and IFIC, Centro Mixto Universidad de Valencia-CSIC. Facultad de Fisica, Universidad de Valencia, Burjassot-46100, Valencia, Spain.}

\begin{abstract}

We explore static and spherically symmetric  solutions of the 4-dimensional semiclassical Einstein's equations using the quantum vacuum polarization of a conformal  field as a source. 
These solutions may be of interest for \black{the study of}  exotic, compact objects (ECOs).
The full backreaction problem is addressed by solving  the semiclassical Tolman-Oppenheimer-Volkoff (TOV) equations making use of effective equations of state inspired \black{by} the trace anomaly and an extra   simplifying and reasonable assumption. 
We combine  analytical and numerical techniques to solve the resulting differential equations, both perturbatively and nonperturbatively in $\hbar$. In all cases the solution is similar to the  Schwarzschild metric up to the vicinity of the classical horizon $r=2M$. However,
 at $r=2M + \varepsilon$, with $\varepsilon\sim O(\sqrt{\hbar})$, 
we find  a coordinate singularity. In \black{the} case of matching with a static star, this  leads to an upper bound in the compactness,
and sets a constraint on the family of stable ECOs. We also study the corrections that the quantum-vacuum polarization induces on the propagation of waves, and discuss the implications.
For the pure vacuum case,  we can further extend the solution by using appropriate coordinates  until we reach  another singular point, 
where this time a null  curvature singularity arises and prevents extending beyond. 
This picture qualitatively agrees with the results obtained in the effective two-dimensional approach, and reinforces the latter as a reasonable method.
 \end{abstract}

\maketitle

\section{Introduction}

Advances in gravitational wave (GW) astronomy to detect and analyze GWs in the last years \cite{LIGO-Virgo}, as well as the recent progress in very long baseline interferometry \cite{EHT}, are opening new avenues to study strong-field gravity and the physics of black holes.  In particular, with the advent of large amounts of data from GW and electromagnetic observations in the future, it will become possible to test and to quantify in precise terms the existence of horizons. 
As a result, there is \black{a growing} interest in studying  models of  dark, compact horizonless astrophysical \black{objects that} may mimic very closely the behavior of  black holes in the GW data, and \black{in examinating} different physical mechanisms that could be used to  uncover these exotic compact objects (ECOs)  with observations \cite{Cardoso-Pani-review}.

While there exists a large class of different models that manage to simulate black holes, most of them require going beyond the Standard Model of particles and/or general relativity (GR) \cite{Shapiro, Schunck, Cardoso, Mazur-Mottola, Holdom, Mathur, Brunstein}. This is because similar values of BH compactness are required to mimic GW observations, but stable astrophysical objects with such compactness are forbidden within GR  by Buchdahl's theorem and the classical energy conditions.  
An appealing possibility is to consider   quantum effects (\black{while preserving} classical gravity as described by conventional GR), as they can potentially avoid the assumptions of this theorem without requiring exotic assumptions. This involves facing  the difficulties of the renormalized stress-energy tensor $\langle T_{ab} \rangle$, describing \black{the} gravitational vacuum polarization of quantum fields, and also solving the corresponding semiclassical backreaction equations. So far all  methods developed  to compute $\langle T_{ab} \rangle$ in quantum field theory in curved spacetime, either analytical or numerical, assume a fixed background metric. Even fixing the background, the explicit computation of $\langle T_{ab} \rangle$ is complicated and only a few examples are known, 
mainly in cosmology \cite{birrel-davies, parker-toms, hu-verdaguer} and for stationary configurations  \cite{Anderson95, Levi-Ori, Levi-Ori-2}. 
As a consequence, the problem of solving the full semiclassical Einstein's equations is terribly complicated, even approximately. Since the nontrivial $(t-r)$ part of a spherically symmetric metric is two-dimensional, a popular approach in the past \black{has been} to consider the analogous problem  in effective $1+1$ dimensions.  A first attempt in this direction  is to truncate the theory to  the $s$-wave sector of the matter field and implement dimensional reduction by integrating the angular degrees of freedom. One ends up with an effective two-dimensional  theory (i.e., a particular dilaton-gravity theory \cite{FBI-Fabbri}), which, after further simplifying assumptions (near-horizon approximation), has a semiclassical description univocally determined by the two-dimensional trace anomaly $\langle T \rangle = \frac{\hbar}{24\pi} R^{(2)}$ (this is usually \black{referred to} as the Polyakov theory approximation \cite{FBI-Fabbri}). 
In two-dimensions the trace anomaly \black{is sufficient} to fix the quantum stress-energy tensor, which in turn can be used to produce a reasonable approximation \black{for evaluating} static quantum corrections to  the  Schwarzschild geometry in vacuum. The semiclassical solution is similar to the classical Schwarzschild solution until very close to the event horizon, but the near-horizon geometry is replaced by a  bouncing surface for the radial coordinate, mimicking the  throat of a nonsymmetric wormhole. A   curvature singularity is found beyond the  throat  \cite{Fabbri-Navarro}. This picture has been confirmed with more \black{analytical} details in \cite{Ho-Matsuo18} and also in \cite{Julio} (using a natural deformation of the Polyakov theory approximation),  
and very interesting extensions for stellar configurations \black{have} been analyzed in \cite{Carballo-Rubio, Ho-MatsuoJHEP18, Julio-2}.

The above two-dimensional effective method is  expected to \black{provide} important insights, but since the problem is very relevant and it is not entirely clear to what extent the two-dimensional approach is really a good approximation, an intrinsic four-dimensional approach is demanded. This is one of the aims of this work.  Our strategy here will be to solve the full semiclassical Einstein's equations but without explicitly \black{calculating}  $\langle T_{ab} \rangle$. Instead, we shall approach the problem as in classical general relativity, by simply giving equations of state and some appropriate boundary conditions.  One of the equations of state will be determined by the four-dimensional trace anomaly, which is independent of the choice of quantum state. More specifically, we will consider a conformal  quantum field, in which the trace of $\langle T_{ab} \rangle$ is entirely determined \black{by} the anomaly. 
Then, we will assume a natural condition on the tangential pressure which we expect \black{to} capture the main qualitative aspects of the actual solution (we differ here from the assumptions given in \cite{ho-kawai-matsou-yokokura}).
This will make the problem manageable  and will allow us to approach the problem directly in four dimensions.  

In this new framework we will also be interested in \black{investigating}  whether there exists physically reasonable, horizonless ``vacuum'' geometries which may mimic black holes (e.g. wormholes), as well as \black{analyzing} what implications the quantum vacuum-polarization  from the exterior geometry may have on static ECOs.  Uniqueness theorems in classical GR \black{tell} us that the exterior vacuum solution of any ECO must be described by the Schwarzschild metric, and this is  widely \black{taken} for granted  in the literature. However,  quantum fields exist all around, and their presence may break this degeneracy with respect to black holes.

Even though semiclassical gravity may provide a conservative framework for studying the formation and/or exterior geometry of exotic astrophysical objects, for solar-mass scales
it is often expected that quantum effects should only lead to extremely low corrections of \black{the} classical solutions, in such a way that from an observational point of view the difference is  totally negligible. Remarkably, recent works developed by different independent groups have shown that even tiny corrections to the metric may \black{significantly alter} the quasinormal mode (QNM) frequency spectrum of black holes \cite{Jaramillo-Macedo, Cardoso-Jaramillo, ACDRMP, Chakraborty}, opening the possibility of constraining these quantum corrections with GW spectroscopy.  Incidentally, this provides a fantastic opportunity to test quantum field theory in astrophysics
and adds further motivation to address the historical difficulties encountered when solving the semiclassical Einstein's equations.

The paper is organized as follows. In Sec. II we provide the setup of the calculation by writing down the central equations, 
as well as by specifying and motivating the assumptions in our problem. Then in Sec. III we solve the differential equations, combining both analytical and numerical techniques, and highlight the main features of the solution obtained, as well as the implications for ECOs. 
In Sec. IV we obtain the maximal extension and describe the curvature singularity that arises.
Sec. V is devoted to physical applications of the \black{obtained} semiclassical metric. In particular we derive the dynamical equations governing scalar and electromagnetic waves,
estimate the associated  light-ring frequencies  
using the WKB approximation, 
and compare \black{them} with the Schwarzschild case.
Finally, in Sec. VI we present our conclusions. 

Our conventions are as follows. We work in  geometrized units $G=c=1$ and keep $\hbar$ explicit \black{throughout}. The metric signature has signature $(-,+,+,+)$, $\nabla_a$ will denote the associated Levi-Civita connection, the Riemann tensor is defined by\, $2 \nabla_{[a}\,\nabla_{b]}v_c=:R_{abc}{}^{d}v_d$\, for any 1-form $v_d$; the Ricci tensor is  defined by $R_{ab}:=R_{acb}{}^c$; and the scalar curvature is $R:=g^{ab}R_{ab}$. All tensors and functions  are assumed to be smooth, unless otherwise stated. \\

\section{Semiclassical TOV equations in quantum vacuum} \label{semi_TOV}

Our aim in this work is to study solutions of the semiclassical Einstein's equations 
\be \label{Eseq}G_{ab} = 8\pi (\langle T_{ab} \rangle + T^{classical}_{ab})\, , \ee
in order to find an effective metric that may describe quantum corrections to  classical black hole spacetimes induced by the quantum vacuum, or even a new family of solutions. Here $T^{classical}_{ab}$ represents some classical gravitational source, while $\left<T_{ab}\right>$ denotes the expectation value of the stress-energy tensor, evaluated for some vacuum state $\left|0\right>$ of some given  quantum field living on the background metric $g_{ab}$ that solves the above equations. For $ T^{classical}_{ab}=0$ and in the absence of quantum fields the spherically symmetric solution is a Schwarzschild black hole due to  Birkhoff's theorem. But if a quantum field is included, $\langle T_{ab} \rangle\neq 0$, and
we expect to get a Schwarzschild-type deformed metric due to quantum vacuum effects ascribed to that field. Solving this problem requires finding a vacuum state $\left|0\right>$ and a metric $g_{ab}$ that together solve (1). For reasons that we \black{will discuss in more detail}  below, this is  an extraordinary problem and there are \black{currently} no systematic techniques available to address the full question. Our strategy will consist in fixing some desirable properties for the vacuum state and solving the resulting PDE for $g_{ab}$.  
More precisely, we will demand the vacuum state to be static and invariant under the group of rotations.  This may be thought of as the most immediate quantum generalization of the classical Schwarzschild vacuum. The solution to (1) will \black{then} correspond to a spherically symmetric and static metric, which in global coordinates $\{t,r,\theta,\phi\}$ can be written  as \cite{Schutz}
\be ds^2 = -e^{-2\phi(r)} dt^2 + \frac{dr^2}{1-\frac{2m(r)}{r}} +r^2 d\Omega^2 \, . \label{schwansatz}\ee
Physically, the assumption of staticity is fundamental \black{for studying}  the exterior vacuum region of exotic compact objects (ECOs) that are stable. For black holes, on the other hand, it is well-known that the assumption of staticity  leads to the Boulware state, which gives rise  to divergences in the stress-energy tensor at the classical horizon \cite{candelas}. However,  this conclusion holds \black{only}  when the renormalized stress-energy tensor is computed for a test quantum field on a fixed Schwarzschild background. In this work we will evaluate the implications of staticity when \black{considering the whole}  problem, including the backreaction effect that the quantum vacuum may produce in the metric.

To get the specific values of the metric components in (2) we have to solve (1) for $T_{ab}^{classical}=0$. 
For a static and spherically symmetric vacuum state the most general expression for the renormalized stress-energy tensor $\langle T_{ab} \rangle$ is
\bea
 \langle T_{ab} \rangle =  -\langle \rho (r)\rangle u_a u_b +  \langle p_r (r)\rangle r_a r_b + \langle p_t (r)\rangle q_{ab}\, ,
\eea
where $u_a= e^{-\phi}\nabla_a t$ is a timelike vector normalized as $u^2 =-1$,  $r_a=(1-\frac{2m(r)}{r})^{-1/2}\nabla_a r$ is a unit spacelike vector, and $q_{ab}$ is the metric on the unit 2-sphere. The metric can be written covariantly as $g_{ab}=-u_a u_b + r_a r_b+ q_{ab} $. There are only three independent equations from the semiclassical Einstein equations.
On the other hand,  there is one nontrivial Bianchi identity. 
Collecting the $tt$ and $rr$ Einstein's equations and this Bianchi identity we get the following equations
\bea 
\frac{d m(r)}{dr}&=& 4\pi r^2\langle \rho (r)\rangle \, , \label{eqnmass}\\
\frac{d \phi(r)}{dr}&=&-\frac{m(r) +4\pi r^3 \langle p_r (r)\rangle }{r^2(1- \frac{2m(r)}{r})} \label{eqnphi} \, ,\\ 
\frac{d \langle p_r (r) \rangle }{dr}&=&-\frac{m(r)+4\pi r^3\langle p_r(r)\rangle }{r^2(1- \frac{2m(r)}{r})}(\langle \rho(r) \rangle  +\langle p_r(r) \rangle )-\frac{2}{r}(\langle p_r(r) \rangle  -\langle p_t (r) \rangle) \ .  \label{eqpressure}
\eea
When $\langle p_r \rangle  \neq \langle p_t \rangle $,  there are anisotropic pressures. In the isotropic case this system of equations reduces to the usual Tolman-Oppenheimer-Volkoff (TOV) equations. 
In the rest of the work we will \black{refer to}  this system of equations \black{as}  the semiclassical TOV equations. 

In this system there are 5 unknowns (3 from the stress-energy tensor and 2 from the metric) for 3 equations. \black{Normally} one would compute $\left<T_{ab}\right>$ and express the result in terms of $\phi(r)$ and $m(r)$ in order to get the system above solved. Instead, we will impose  two  functional relations between the components of the stress-energy tensor, in order to avoid such a difficult (or unattainable) calculation. First, we 
will consider the case of  a  massless quantum field conformally coupled to the spacetime.  The advantage of doing this  is that the relation between the three independent components of the stress energy tensor is univocally fixed by the trace anomaly $\langle T^a_a \rangle$ as
\be -\langle \rho \rangle +\langle p_r \rangle +2\langle p_t \rangle= \langle T^a_a \rangle\, ,  \label{stateeq}\ee
and the trace anomaly  is uniquely determined by the geometry of the spacetime
\bea \langle T_a^a \rangle = \frac{\hbar}{2880 \pi^2}( \alpha C^{abcd}C_{abcd} + \beta R^{ab}R_{ab} + \gamma R^2 +\delta \Box R) \ .  \label{trT} \eea
 In this expression $C_{abcd}$ is the Weyl tensor, $R_{ab}$ the Ricci tensor, $R$ the Ricci scalar and $\alpha,\beta,\gamma,\delta$ are  real numbers.  Most importantly, this result is independent of the choice of the quantum state. The idea of exploiting the trace anomaly goes back to \cite{Christensen-Fulling}.
The constant coefficients depend on the particular field under consideration. 
It should be noted though that there exits an intrinsic  ambiguity in the trace anomaly for the coefficient $\delta$ \cite{Wald78}. This ambiguity is related to the choice of  the renormalization scheme. 
The term with $\Box R$ can always be removed by adding a local counterterm in the Lagrangian so, \black{from now on} we set $\delta=0$.  This simplifies the problem \black{considerably}, since it will avoid derivatives of second and third order of the metric in the field equations. 

By evaluating (\ref{trT}) with our metric and using the semiclassical TOV equations written above 
one can obtain a simplified expression for the trace anomaly in terms of $\langle\rho\rangle$, $\langle p_r\rangle$ and $\langle p_t\rangle$. This leads to the following equation of state
\be
-\langle\rho\rangle+\langle p_r\rangle+2\langle p_t\rangle=\frac{\hbar}{270}\left[\alpha\left(\frac{3 m}{ 4\pi  r^3}-\langle\rho\rangle+\langle p_r\rangle-\langle p_t\rangle \right)^2+6 \beta \left(\langle\rho\rangle^2+\langle p_r\rangle^2+2\langle p_t\rangle^2\right)+6 \gamma (-\langle\rho\rangle+\langle p_r\rangle+2\langle p_t\rangle)^2\right]\, .
\ee
For definiteness in this work we restrict to  scalar fields, for which the coefficients are $\alpha=\beta=1$ and $\gamma=-1/3$. For these values the above expression can be further simplified to
\be
-\langle\rho\rangle+\langle p_r\rangle+2\langle p_t\rangle=\frac{\hbar}{180}\left[\frac{m}{\pi r^{3}}\left(3 \frac{m}{\pi r^{3}}+8( -\langle\rho\rangle+\langle p_r\rangle-\langle p_t\rangle)\right)+8 \langle\rho\rangle\left(\langle\rho\rangle-\langle p_r\rangle+2\langle p_t\rangle\right)+8(\langle p_r\rangle-\langle p_t\rangle)^2\right]\, . \label{eqnstate}
\ee

We need another restriction to make our system of equations solvable. Unfortunately there are no other universal geometric properties of the stress-energy tensor that may allow us to fix a similar relation between the different components of the stress-energy tensor.  
To proceed further we need to impose a  condition on $\langle T_{ab}\rangle$ based on what we may expect from the quantum state. 
We will consider here that $\langle p_r \rangle = \langle p_t \rangle$. This simplifying assumption is inspired \black{by}  the ``zero-order'' result that one gets when calculating $\langle T_{ab}\rangle$  in a fixed Schwarzschild background when $r\to 2M$, and we expect this near-horizon approximation to capture the qualitative behavior of the actual solution.
Indeed, in a Schwarzschild spacetime background  the vacuum expectation value $\langle T_{ab} \rangle$ of a  conformal scalar field in the  static spherically symmetric  state  behaves, in the vicinity  of the horizon, as \cite{candelas} 
\be\label{Candelas}
\left\langle T_{\mu}^{\nu} \right\rangle
\sim- \frac{\hbar}{2 \pi^{2}(1-2 M / r)^{2}}
\ \int_{0}^{\infty} \frac{d \omega \omega^{3}}{e^{8 \pi M \omega}-1}\left[\begin{array}{cccc}
-1 & 0 & 0 & 0 \\
0 & \frac{1}{3} & 0 & 0 \\
0 & 0 & \frac{1}{3} & 0 \\
0 & 0 & 0 & \frac{1}{3}
\end{array}\right] \, .
\ee
Both $\langle T^\theta_\theta \rangle \equiv \langle p_t\rangle$ and $\langle p_r \rangle\equiv \langle  T^r_r\rangle$   merge for $r \to 2M$, but \black{as one moves away}  from the vicinity of the horizon, the tangential and radial pressures start \black{to differ}.  In fact, for $r \to \infty$ one has $\langle p_r\rangle = -\frac{1}{3} \langle p_t \rangle \sim \mathcal O(r^{-5})$ \cite{Anderson-Balbinot-Fabbri}. Therefore, our assumption is expected to work only qualitatively as an approximation to the actual relationship, whose knowledge requires computing $\langle T_{ab} \rangle$  in detail. 
This simplification  is expected to capture the main physical ingredients of our field theory  (the results obtained will  be exact \black{at least}  in a neighborhood of the classical horizon).

Our approach can be easily compared \black{with}  other works by fixing this free condition with different assumptions. For instance, the effective two-dimensional Polyakov approximation \cite{FBI-Fabbri, Fabbri-Navarro, Ho-Matsuo18, Ho-MatsuoJHEP18, Carballo-Rubio} can be regarded as fixing trivially the tangential pressure $\langle p_t\rangle=0$ (or with additional extra deformations \cite{Julio, Julio-2}) and restricting the trace anomaly to its  two-dimensional value.  
\black{Instead, we are}  trying to solve  the 4D problem directly without assuming {\it a priori} that it is   similar to the 2-dimensional case. 
 On the other hand, the approach of \cite{ho-kawai-matsou-yokokura}  also quantizes the matter field in four dimensions, but assumes that $\langle p_t \rangle$ is regular as $r\to 2M$, even in the Schwarzschild background. Instead, our assumption is compatible with Eq. (\ref{Candelas}).

\section{Semiclassical metric solution} \label{solution}

\subsection{Perturbative analytical solution} 

 The leading order contributions of the stress-energy tensor are expected to behave as $\langle\rho\rangle\sim O(\hbar^1)$, $\langle p\rangle\sim O(\hbar^1)$ [where $\langle p\rangle=\langle p_r\rangle=\langle p_t\rangle$]. We can thus look for perturbative solutions of the semiclassical TOV equations, solving the system order by order in powers of $\hbar$.  In this subsection we will obtain the first order correction using analytical techniques, and in the next subsection we  will analyze the validity of this approach by  solving the system of equations numerically. 

Solving the TOV equations at order $\hbar^0$ gives  $m(r)=M+O(\hbar)$ and $\phi\sim-\frac12\log ({1-2M/r})+O(\hbar)$, where $M$ is an arbitrary constant of integration, which can be identified with the ADM mass. This is the Schwarzschild metric, as expected at order $\hbar^0$. 
To get something interesting we have to solve the equations at first order in $\hbar$.  Let us define $m=M+m_1 \hbar+O(\hbar^2)$, $\phi\sim-\frac12\log ({1-2M/r})+\phi_1 \hbar +O(\hbar^2)$, $\langle \rho\rangle=\rho_1 \hbar+O(\hbar^2)$,$\langle p\rangle=p_1\hbar+O(\hbar^2)$. Then the system of equations at first order in $\hbar$ is given by
\bea 
\frac{d m_1}{dr}&=& 4\pi r^2\rho_1 \, ,\\
\frac{d \phi_1}{dr}&=&-\frac{m_1}{r^2 f^2}-\frac{4 \pi  r p_1}{f}\, , \\ 
\frac{d p_1 }{dr}&=&-\frac{M}{r^2f}(\rho_1+p_1)\, , \\
-\rho_1&+&3p_1= \frac{ M^2}{60\pi^2 r^{6}} \, ,
\eea
where $f=1-\frac{2M}{r}$. This system can be solved analytically, obtaining the following expressions for the pressure and density
\footnote{The negative sign and the dependence on $1/f^2$ obtained in these expressions \black{are} in agreement with the exact  results obtained on the fixed (Schwarzschild) background near the horizon for the Boulware vacuum state [see (\ref{Candelas})].}
\bea
\langle p\rangle&=& -\frac{\hbar M^3}{480\pi^2r^7 f^2}(\frac{1}{7} +f) + \mathcal{O}(\hbar^2)\ , \label{p1}\\
\langle\rho\rangle &=& \hbar\left( -\frac{M^3}{160 \pi^2 r^7 f^2}(\frac{1}{7} +f) - \frac{ M^2}{60 \pi^2 r^6} \right)+ \mathcal{O}(\hbar^2)\ , \label{rho1}
\eea
and the following ones for the metric components
\bea
m&=&M+\frac{\hbar}{40320 \pi  Mf }\left(9-36 f \log (f)+10 f-174 f^2+246 f^3-91 f^4\right)+O(\hbar^2)\, , \label{m_pert}\\
\phi&=&-\frac12\log f+\frac{\hbar}{80640 \pi  M^2f^2}\left(3+36 (-1+3 f) f \log (f)-35 f+152 f^2-132 f^3+5 f^4+7 f^5\right)+O(\hbar^2)\, . \label{phi_pert}
\eea   
 To fix the constants of integration we have assumed the natural boundary conditions $\langle p\rangle(r\to\infty) =0$, $\langle \rho\rangle(r\to\infty) =0$ and the metric tending to the Schwarzschild one as $r\to\infty$.
For pedagogical purposes, we display the asymptotic form of the metric around $r=2M$ 
\be
ds^2=-\left(f(r)-\hbar\left(\frac{1}{13440\pi M^2 f(r)}+\mathcal{O}(\log f)\right)+\mathcal{O}(\hbar^2)\right) dt^2 + \frac{dr^2}{f(r)-\hbar\left(\frac{1}{4480\pi M^2 f(r)}+\mathcal{O}(\log f)\right)+\mathcal{O}(\hbar^2)} + r^2 d\Omega^2 \, . \label{analyticalresult}
\ee
In the Appendix \ref{appendix} we prove that   the curvature  at this singular point \black{is finite}, so this  is just a coordinate singularity. In fact, this is  just  the classical Schwarzschild coordinate singularity at $r= 2M$ shifted to the value $r_0$ defined by $g_{rr}^{-1}(r_0)=0$. Using the expression (\ref{m_pert}) and imposing $2m(r_0)=r_0$, 
we easily obtain
\be r_0 = 2M + \frac{\sqrt{\hbar}}{4\sqrt{70\pi}} +\mathcal{O}(\hbar) \ . \label{singularpoint}\ee
In geometrized  units $\sqrt{\hbar}=l_p$ is the Planck length. This singular, limiting point defines the end of validity of our coordinate system, which  would traditionally indicate the location of a ``horizon'' at $r=r_0$. However, note that, unlike the Schwarzschild case, in this point the component $g_{tt}$ of the metric (the so called redshift function) does not vanish, but takes the value
\be \label{redshift}
g_{tt}(r_0)= -\frac{\sqrt{\hbar}}{12\sqrt{70\pi}M}+\mathcal{O}(\hbar)\, .\ee
This implies that the static spacetime that we have obtained does not contain a horizon, i.e. it is not defining a black hole \cite{Vishveshwara}. We check this in the Appendix \ref{appendix}.

Note that, even though   (\ref{p1})-(\ref{rho1}) are generally  very small (because of the prefactor $\hbar$), they  become relevant around  $r\sim r_0$, since  in this limit the factor $f(r)$ in the denominator can compensate $\hbar$. In other words, quantum effects are quite important near the location of what \black{was} classically  the horizon. The Krechtmann scalar is also found to \black{be} significantly corrected at the singular point (see Appendix \ref{appendix}). These observations  \black{lead} us to the following subsection.

\subsection{Nonperturbative numerical solution} 

As we can see  the results obtained above at first order in $\hbar$ \black{also depend}  on $f(r)$, which takes values of order $\sqrt{\hbar}$ near the singular point $r=r_0$. This \black{dependence}  compensates the small value of $\hbar$ in some expressions above. Because of this, a natural question is whether the perturbative method is a good approximation near to the singular point. To answer this we can solve the TOV equations at second order in $\hbar$ and analyze whether  near the singular point the solution is consistent with the perturbative hypothesis (i.e. that the order $\hbar^1$ is larger than the  order $\hbar^2$, etc.). 
The analysis is tedious and we avoid showing the details. What we obtain is that the  $\hbar^2$ contribution to    the pressure and the density \black{is} proportional to $\hbar^2/{f(r)^4}$. Near to the singular point $f(r)^4$ is of order $\hbar^2$, so this term competes with  the first order contribution \eqref{p1}, which is proportional to $\hbar/f(r)^2$. Therefore we find that, near the singular point, the higher order contributions in $\hbar$ are not necessarily smaller than the first one and perturbation theory actually breaks down. Therefore, \black{we} cannot rely on the perturbative series in the \black{vicinity of} $r_0$  
and we \black{are forced}  to solve the differential TOV equations exactly, which can only be done numerically. Still, we shall find that the perturbative approach presented in the previous subsection is a good approximation to the problem, and  it \black{qualitatively}  predicts well the behavior of the nonperturbative solution.\footnote{ In this paper we work in the semiclassical regime in which fluctuations of the stress-energy tensor are negligible  compared to its mean value. Going beyond this framework would require working with techniques in stochastic gravity \cite{hu-verdaguer}, which is out of the scope of the present paper. By nonperturbative we mean the exact solution of \black{the} TOV equations within the \black{semiclassical} framework.}

We  \black{now turn} to solve numerically the TOV equations (\ref{eqnmass})-(\ref{eqpressure}) using the  equation of state (\ref{eqnstate}) and $\langle p_r \rangle  = \langle p_t \rangle $. We place the boundary conditions at $r=1000M$, and demand that \black{at} this location the solution is approximately the Schwarzschild metric\footnote{To get more precision we can choose the corrected solution at first order in $\hbar$ obtained above, but the results near the singular point are numerically indistinguishable.}. 
An important issue that one faces when  solving  the equations numerically is that the value of $\hbar$ is much smaller than $M$. To be able to distinguish the implications of a nonzero but tiny value of $\hbar$ from the numerical error, one needs a huge computer accuracy.  
To avoid this issue, a useful strategy  is to use first some artificial high values of $\hbar$ (between $10^{-5}M^2$ and $10^{-15}M^2$), study the dependence of the results \black{on} $\hbar$, and then extrapolate the relevant quantities to the actual value of $\hbar$. \black{By} solving numerically the equations for different values of $\hbar$ and calculating for each case the value of $r_0$ we obtain results that \black{approximately} fit  the expression $r_0\approx 2M+0.01947\sqrt{\hbar}$. This shift differs from the one  estimated \black{by} the perturbative method
($r_0\approx 2M+0.01686\sqrt{\hbar}$) but the functional dependence on  $\hbar$ remains the same.

\begin{figure}[htbp]
\begin{center}
\includegraphics[width=85mm]{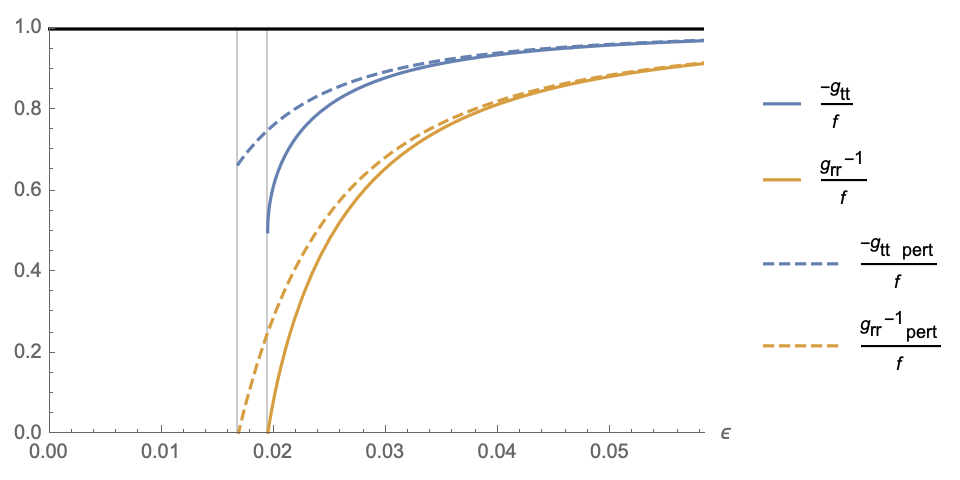}
\includegraphics[width=85mm]{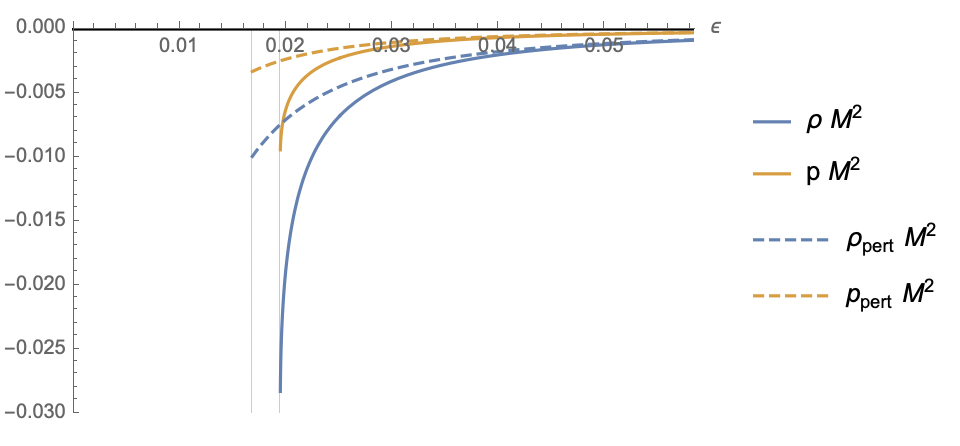}
\end{center}
\caption{Numerical results obtained for the metric components and the renormalized energy density and pressure near the singular point $\epsilon=0.01949$ (where $r=2M+\epsilon\sqrt{\hbar}$). We have chosen $\hbar/M^2=10^{-5}$, but the plots do not change significantly for other values. We compare them with the perturbative solution (dashed curves), for which the singular point is $\epsilon=0.01686$.
}
\label{corrections}
\end{figure}

In Fig. \ref{corrections} we \black{plot}  the components of the metric obtained numerically, normalized by the factor $f(r)=1-2M/r$, as well as the renormalized energy density and pressure. They are plotted  as a function  of  $\epsilon=\frac{r-2M}{\sqrt{\hbar}}$. With this new radial variable the singular point $r_0$ does not depend on the specific value of $\hbar$. These plots are taken for $\hbar/M^2=10^{-5}$, but we have analyzed them for other values and have seen that they do not  significantly depend on the chosen value of $\hbar$ near the singular point.  From these plots one can see that, as in the perturbative solution, the component $g_{rr}^{-1}$ tends to 0 at the singular point $r=r_0$, while $g_{tt}$ tends to a nonzero value. The energy density and pressure differ from 0  as they approach the singular point, as expected.
More precisely $g_{tt}\sim O(\sqrt{\hbar}/M)$, $g^{-1}_{rr}\sim (r-r_0)/M$,  $\rho\sim O(\hbar^0)$, and $p\sim O(\hbar^0)$ as $r\to r_0$. This is the same dependence on $\sqrt{\hbar}/M$ as that obtained by the perturbative approach, although the numerical coefficients \black{are different}. This allows us to consider the perturbative solution as a qualitatively good approximation. 

We can summarize the above numerical result in terms of the following generic expression for the metric 

\be 
ds^2=g_{tt}dt^2 + g_{rr}dr^2+ r^2 d\Omega^2 \label{numericalresult}
\ , \ee
where $g^{-1}_{rr} \to 0 $, as $r \to r_0 >2M$ and $g_{tt}(r_0)\neq 0$. Furthermore, $g^{-1}_{rr} \ \sim (r-r_0)/M$ and    $g_{tt}(r) \sim \mathcal{O}(\sqrt{\hbar}/{M})$ in a neighborhood of $r_0$.

\section{Extension beyond the coordinate singularity.}\label{extension}

The metric (\ref{numericalresult}) (or (\ref{analyticalresult})) is only meaningful when $r>r_0$ because of the coordinate singularity at $r=r_0$. 
We recall (see Appendix) that \black{the} curvature scalars are finite at $r=r_0$. 
Physically this effective metric can be used to describe the exterior spacetime of \black{a} static, spherically symmetric star, including the vacuum polarization effects  of quantum fields around.  But in close analogy to the classical Schwarzschild case when expressed in $\{t,r,\theta,\phi\}$ coordinates, one may attempt to extend the spacetime across the $r=r_0$ point and examine if there exists a purely (quantum) vacuum solution.  As remarked at the end of Sec. III.A, the usual Eddington-Finkelstein coordinates fail to provide a regular metric, which prevents the usual analytical extension beyond $r=r_0$. 

By looking at the specific form of the metrics (\ref{numericalresult}) or (\ref{analyticalresult})  one realizes that they can be used to construct a portion of a  static,  traversable  (and Lorentzian) wormhole \cite{Visser, Lobo}.  By introducing the usual proper-length coordinate $l(r)\equiv \int_{r_0}^{r} 1/\sqrt{1-2m(r')/r'}dr'\geq 0$ the metric can be rewritten to fit  the Morris-Thorne ansatz 
\bea
ds^2=-e^{-2\phi(l)}dt^2+dl^2+r(l)^2d\Omega^2\, . \label{MTansatz}
\eea
Therefore,  one can  extend the spacetime  beyond the critical point $r=r_0$ or $l=0$ (\black{which physically represents} the throat of the wormhole) by analytically extending to negative values of $l$. 
The function $r=r(l)$ is determined by inverting the equation $l=l(r)$ given above, but only when $l>0$.  For $l<0$ the function $r=r(l)$ must be determined by other means.

\subsection{Setup}

Instead of working with the metric ansatz (\ref{schwansatz}) and then \black{transforming} to (\ref{MTansatz}) by a change of variables, we can alternatively solve the  problem from scratch using the latter metric directly and explore if there exist wormhole solutions. The equivalent system of TOV equations now \black{reads} (we find convenient to introduce the defining relation $g(l) \equiv \frac{dr}{dl}$)
\bea
\frac{d r}{dl} & = & g \, ,\label{eqn1}\\
\frac{dg}{dl} & = & \frac{1-8\pi r^2 \left<\rho\right> +g^2}{2r} \, , \label{eqn2}\\
\frac{d\phi}{dl} & = & \frac{-1-8\pi r^2 \left<p_r\right> +g^2}{2r g} \, , \label{eqn3}\\
\frac{d\left<p_r\right>}{dl} & = &  \frac{(-1-8\pi r^2 \left<p_r\right> +g^2)(\left<p_r\right>+\left<\rho\right>)}{2r g} + \frac{2g(\left<p_t\right>-\left<p_r\right>)}{r}  \, .\label{eqn4}
\eea
There are six unknowns for four equations. Again, we can impose two equations of state to get a solvable model. As before, we shall take $\left<p_t\right>=\left<p_r\right>$ (notice that the contribution of $\langle p_t\rangle-\langle p_r\rangle$ is negligible near the throat, where as we will see $g(0)=0$) and $\left<T^a_a\right>$ given by the trace anomaly:
\be \label{eqstate}
-\langle\rho\rangle+3\langle p\rangle=\frac{\hbar}{180}\left[\frac{1-g^2}{2\pi r^{2}}\left(3 \frac{1-g^2}{2\pi r^{2}}-8 \langle\rho\rangle\right)+8 \langle\rho\rangle(\langle p\rangle+\langle\rho\rangle)\right]\, . 
\ee

To get wormhole solutions we must impose  several  conditions. Without loss of generality, we can locate the throat at $l=0$. One of the sectors of the throat (that would represent the universe we live in) must be asymptotically flat, and inertial observers at infinity must measure time with $t$. We choose that sector  corresponding to $l>0$. Then the previous condition requires $\phi(\infty)=0$, $\left<p_r\right>(\infty)=\left<\rho\right>(\infty)=0$. On the other hand,  the coordinate $l$ should agree with the radial function $r(l)$ at infinity, i.e. $r(l) \to l$ as $l\to \infty$. Furthermore,  for sufficiently large distances away from the throat, $g$ must be given by the Morris-Thorne coordinate transformation (the solution should mimic a black hole at large distances), i.e. $g(l)\sim \sqrt{1-2m(l)/l}$ and therefore $g(\infty)=1$. 

This set of boundary conditions, together with the two equations of state specified above, can be used to obtain a unique solution to the above system of differential equations, integrating all the way down from $l=+\infty$  until negative values of $l$. Notice that in general  there will be \black{no} mirror-reflection symmetry at the throat. The results are shown in the next subsection.  For the solution to represent a wormhole, note that i) the throat must have a finite, nonvanishing radius, so $r(0)=r_0>0$, and ii)  the throat area must correspond to a minimum, therefore $g(0)=0$. 

Before discussing the results, we remark \black{an} important issue. It may \black{seem} that the \black{above} system of equations is not well defined at the throat $l=0$ because of $g(0)=0$ in the denominator of some equations. But notice that, according to Einstein's equation,
\bea
\left<p_r(r)\right>=-\frac{1}{8 \pi}\left[\frac{2m}{r^{3}}-2\left(1-\frac{2m}{r}\right) \frac{\partial_r\phi}{r}\right]\, ,
\eea
so at the throat (where $2m(r)=r$) we also have $p_r(0)=-1/(8\pi r_0^2)$, provided that $\partial_r \phi(0)$ is well-defined at the throat or that it does not blow up as quickly as $(1-2m/r)^{-1}$  (this is verified in this case, using the numerical solution obtained in the previous section one can see that $\partial_r\phi\sim(1-2m(r)/r)^{-1/2}$ when $r\to r_0$). Therefore, the numerator of (\ref{eqn3}) and (\ref{eqn4}) also vanishes whenever the denominator does, and we have a $0/0$ ambiguity.  To ensure that we can extend the metric across the throat one needs to check first that the limit $l \to 0$ tends to a finite value under the boundary conditions specified above. Numerically we find that near the throat $1+8\pi r^2 p\sim O(l)$ and $g\sim O(l)$, so  we can conclude that the limit of the quotient will be finite. 

\subsection{Results}
\begin{figure}[htbp]
\begin{center}
\begin{tabular}{c}
\includegraphics[width=60mm]{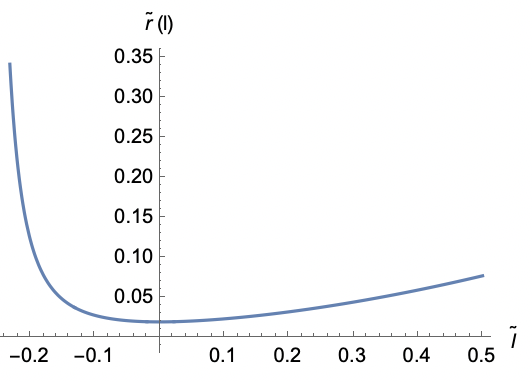}
\includegraphics[width=60mm]{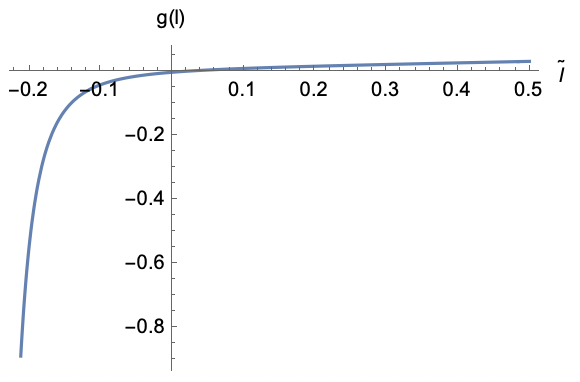}
\includegraphics[width=60mm]{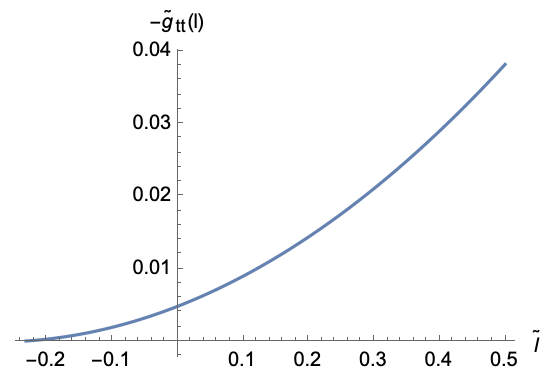}\\
\end{tabular}
\end{center}
\caption{Numerical results obtained for the components of the metric \eqref{MTansatz} and $g(l)=r'(l)$ in terms of $\tilde l=l\, \hbar^{-1/4}M^{-1/2}$. The \black{represented} interval of $\tilde l$ includes the throat ($\tilde l=0$) and the curvature singularity ($\tilde l\approx -0.278$). We have defined the quantities $\tilde r=(r-2M)\hbar^{-1/2}$ and $\tilde g_{tt}=g_{tt}\,\hbar^{-1/2}M$, in such a way that their values at the throat do not depend on the chosen value of $\hbar$. We have chosen $\hbar/M^2=10^{-3}$ for these plots, but they have a similar form for other values. }
\label{singularity}
\end{figure}

In Fig. \ref{singularity} 
we show the result of solving numerically the system of equations \eqref{eqn1},\eqref{eqn2},\eqref{eqn3},\eqref{eqn4} and \eqref{eqstate} under the conditions specified in the previous subsection. As in Sec III. B, to capture the implications of a nonvanishing but tiny value of $\hbar$ on the equations, we do the calculation for several high values of $\hbar$ (so that their effect is numerically distinguishable), then we perform a fit of the results to be able to extrapolate the value of interest with the actual value of Planck's constant. 
In our calculation the throat is located at $l=0$, note how at this point there is a bounce in the function $r(l)$ (its derivative $g(l)$ \black{changes} sign). 

\begin{figure}[htbp]
\begin{center}
\includegraphics[width=90mm]{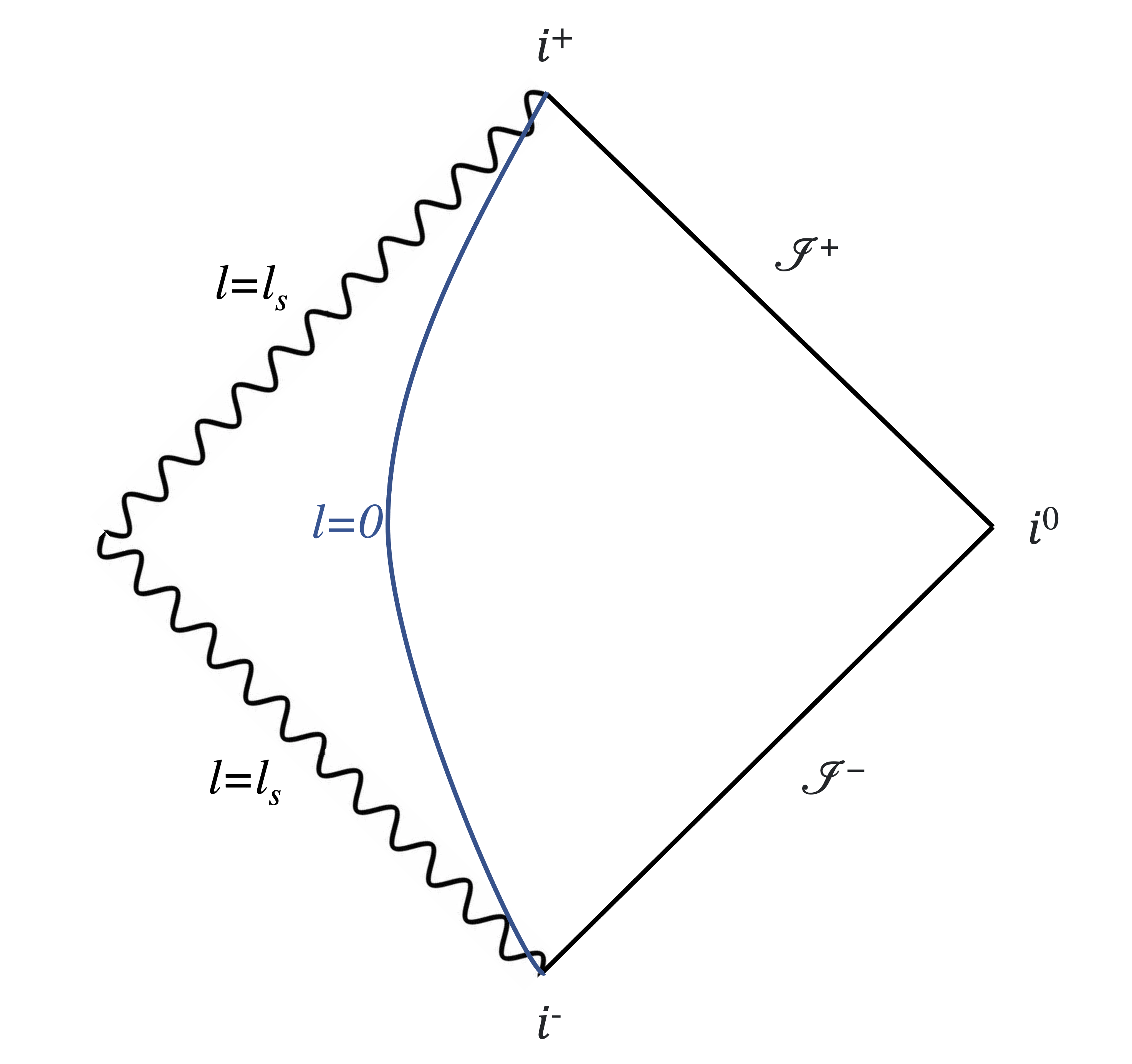}
\end{center}
\caption{Penrose diagram showing the wormhole throat ($l=0$) and the null curvature singularity ($l=l_s$).}
\label{penrose}
\end{figure}

Furthermore, we find that in the interior region, $l<0$, a new singular point appears at $l_s\sim -0.278\, \hbar^{1/4}\sqrt{M}$. It is a singular point because the redshift function vanishes there,  $g_{tt}(l_s)=0$. As we approach to $l_s$  we find that the renormalized density, the  pressure and the scalar of curvature $R=8\pi(-\rho+3p)$ all tend to diverge. 
This signals  the existence of a curvature singularity.
To confirm the existence of this singularity from an analytical viewpoint we can  examine the expression of the scalar curvature in terms of the metric components:
\be
R(l)=\frac{g_{tt}'(l)^2}{2 g_{tt}(l)^2}-\frac{g_{tt}''(l)r(l) +2 g_{tt}'(l)
   r'(l)}{g_{tt}(l) r(l)}-\frac{2 \left(2 r(l) r''(l)+r'(l)^2-1\right)}{r(l)^2}\, .
\ee
Since \black{at} the singular point $g_{tt}(l_s)=0$ (see Fig. \ref{singularity}) some terms of this expression diverge at this point. Although $g_{tt}'(l)$ also vanishes at $l=l_s$, numerical computations show that it decreases slower than $g_{tt}(l)$.  
To see the causal character of this curvature singularity, let us consider the induced metric on a $l=$ constant three-dimensional hypersurface: $d\bar s^2=g_{tt}(l)dt^2+r(l)^2d\Omega^2$. At the singularity $l=l_s$ we have $g_{tt}(l_s)=0$, so  the metric becomes degenerate: $d\bar s^2=0+r(l_s)^2d\Omega^2$.  Therefore, the surface $l=l_s$ becomes a null hypersurface \cite{Galloway}, and  this curvature singularity is null.  Fig. \ref{penrose} provides  a  Penrose diagram that shows all these features.

An important question is how long \black{it would} take for an  observer  \black{crossing} the throat to reach this curvature singularity. To study this let us consider a radial and timelike geodesic  \black{starting at}  $l=0$ (throat) and ending at the singular point $l=l_s$. The relevant geodesic equation for a static and spherically symmetric metric $ds^2=g_{tt}(l)dt^2+g_{ll}(l)dl^2+r(l)^2d\Omega^2$ is given by
\be
\frac{dl}{d\tau}=\pm\sqrt{-g_{ll}^{-1}(E^2g_{tt}^{-1}+\frac{L^2}{r^2}+\mu)}
\ee
where $\tau$ is the proper time, $\mu=+1,0,-1$ for timelike, null and spacelike geodesics respectively, and $E$ and $L$ are constants of motion given by $E=-g_{tt}\frac{dt}{d\tau}$ and $L=r^2\frac{d\phi}{d\tau}$. In our case $g_{ll}=1$, $\mu=1$, $L=0$, and $dl/d\tau<0$ (the geodesic is approaching the singularity), so
\be
\frac{dl}{d\tau}=-\sqrt{E^2g_{tt}(l)^{-1}-1}\, .
\ee
The proper time  needed to reach the curvature singularity from the throat is  \black{then given} by
\be
\Delta\tau=-\int_0^{l_s}\frac{dl}{\sqrt{E^2g_{tt}(l)^{-1}-1}}\, .
\ee
(the condition that the geodesic propagates  \black{into} the future, $\partial_{\tau}t>0$, implies $E^2g_{tt}(l)^{-1}>1$ and guarantees that the integral is real). The order of magnitude of this quantity can be estimated as follows. 
From \eqref{redshift} we know that $g_{tt}(0)\sim \sqrt{\hbar}/M$. Assuming  $E\sim 1$, we have $E^2g_{tt}^{-1}(l)>>1$ in the region of the integration.  \black{Since} $|l_s|\sim\hbar^{1/4}\sqrt{M}<<1$ we can also Taylor expand the integral to finally get
\be
\Delta\tau\sim- \int_0^{l_s} \sqrt{g_{tt}(l)} dl \sim  \sqrt{g_{tt}(0)} |l_s| + O(l_s^2)\sim\sqrt{\hbar}
\ee
So an observer crossing the throat  \black{will almost immediately see} the presence of the curvature singularity.

 Finally, we want to stress that the occurrence of the curvature singularity has been obtained for a purely vacuum semiclassical solution. The presence  of matter producing  very compact stellar objects (ECOs) makes only the outer part of the  solution physically relevant. 
Moreover, these results also suggest a maximum in the compactness of ECOs. This maximum would be given by the minimum of the radial function $r(l)$, i.e. the throat ($r=r_0$). Therefore this maximum of compactness (measured as $2M/r(l)$) is of order 
\be \label{mr}\frac{2M}{r_0} \sim 1- 0.01686\frac{\sqrt{\hbar}}{2M} \ . \ee
We regard (\ref{mr}) as one of the main results of this work. Probing the exterior of the semiclassical metric via scalar and vector perturbations will be the topic of the next section.

 {\bf Remark:} Another way to extend the metric beyond the coordinate singularity $r=r_0$ consists in defining a coordinate $\bar r$ by $\frac{d \bar r}{dr}=e^{-\phi(r)}(1-\frac{2m(r)}{r})^{-1/2}$. In this case the metric has the form
\be
ds^2=-G(\bar r)dt^2+\frac{dr^2}{G(\bar r)}+R(\bar r)^2d\Omega^2\, .
\ee
Using this metric as an ansatz for solving the semiclassical TOV equations we  found that the functions $G$ and $R$ can be  \black{analitically} extended beyond the coordinate singularity $\bar r=\bar r_0$. In particular $R(\bar r)$ reaches a minimum at $\bar r_0$ and starts increasing  for lower values, as expected  \black{for a} wormhole  metric. On the other hand $G(\bar r)$ continues  \black{to decrease} until it reaches the  value $\bar r=\bar r_s$, where $G(\bar r_s)=0$. At this point we  \black{again find} a curvature singularity, which is equivalent to the one found in the other extension explained above. Therefore, with this alternative extension, we obtain the same conclusions. However the approach described above allows  \black{a} higher accuracy in the numerical calculations.

\section{Propagation of waves in the semiclassical metric}\label{perturbations}

The propagation of waves on a given spacetime background provides a way to test  some features of this metric by studying  the  scattering properties of the wave. Furthermore, they provide a means to test the stability of the metric under linear perturbations, which is a necessary condition \black{for} any semiclassical metric that aims to describe  acceptable astrophysical systems. In this section we will study scalar and electromagnetic perturbations around the semiclassical metric constructed in Sec. III. In particular, we will compute the leading order corrections to the light-ring frequency modes. These frequencies depend only on the geometry around the light-ring of the classical black hole, and describe  the early ringdown stage in gravitational wave observations of binary mergers. While the late ringdown stage is expected to be described by the proper QNM frequencies \cite{CFP, KP2017}, the calculation of these \black{is} out of the scope of the present paper.

\subsection{Scalar perturbations}

Let us study the behavior  of a massless scalar field  coupled to a general static and spherically symmetric background, $ds^2=g_{tt}dt^2 + g_{rr}dr^2+ r^2 d\Omega^2$. The  field satisfies the Klein-Gordon equation
\be
(\Box+\xi R)\phi=0\, .
\ee
Since the metric is static and \black{spherically symmetric,} we can look for solutions of the following form 
\be
\phi_{\omega l m}=\frac{1}{r}e^{-i\omega t}Y_{lm}(\theta,\psi)\Psi_{\omega l }(r)\, .
\ee
For a curved spacetime the D'Alembert operator is given by $\Box\phi=\frac{1}{\sqrt{-g}}\partial_\mu(\sqrt{-g}g^{\mu\nu}\partial_\nu\phi)$. The Klein-Gordon equation decouples  and, after some calculations one obtains the following equation for the radial function
\be \label{ODE}
F^2\Psi_{\omega l}''+FF'\Psi_{\omega l}'+(\omega^2-V_l)\Psi_{\omega l}=0\, ,
\ee
where
\bea
&&F(r)=\sqrt{-\frac{g_{tt}}{g_{rr}}}\, ,
\\
&&V(r)=-g_{tt}(r)\left(\frac{l(l+1)}{r^2}-\xi R(r)\right)+\frac{F(r)F'(r)}{r}\, .
\eea
If we define a generalized tortoise coordinate as $\partial_{r^*}=F(r)\partial_r$, then Eq.(\ref{ODE}) can be rewritten in the usual Regge-Wheeler form $\partial_{r_*}^2{\Psi}_{\ell m} +(\omega^2-V){\Psi}_{\ell m}=0$. In particular, for a Schwarzschild background we recover the usual expression.

Now let us study this potential for our particular case, given by $g_{tt}=-e^{-2\phi(r)}$ and $g_{rr}=1-\frac{2m(r)}{r}$, where $\phi(r)$ and $m(r)$ have been obtained in  Sec. \ref{solution}. \black{Using} the semiclassical TOV equations we can \black{rewrite} the potential as
\be
V(r)=e^{-2\phi(r)}\left(\frac{l(l+1)}{r^2}+\frac{2m(r)}{r^3}+4\pi\left((1-6\xi)\langle p(r)\rangle+(-1+2\xi)\langle\rho(r)\rangle\right)\right)\, .
\ee
It is easy to see that in the Schwarzschild limit $\hbar \to 0$ this expression reduces to the usual effective potential for scalar fields. 
We can \black{now use}  the perturbative solution of the corrected Schwarzschild metric
to obtain the correction of the Regge-Wheeler potential at first order in $\hbar$. 
The resulting expression, at first order in $\hbar$, yields 
\bea \label{corrected_pot}
V(r)&=& V_{0}(r)+\frac{\hbar}{20160 \pi  M^2 r^6 f} \biggl[-2688 M^4 \xi  f^2+M r^3 \biggl(- 3 \lambda+(53+32\lambda) f - 40(1+3\lambda) f^2 + 12(-36+\lambda) f^3 \nonumber\\
&+& (664+7\lambda) f^4 - 245 f^5 \biggr)+18 r^4 \left((1 + \lambda)  - (5 + 3 \lambda) f + 4 f\right)f \log (f)\biggr]+O\left(\hbar^2\right)\, , \label{pot_scalar}
\eea
where $f=1-\frac{2M}{r}$, $\lambda=l(l+1)$ and $V_{0}(r)=f(\frac{\lambda}{r^2}+\frac{2M}{r^3})$, which is the effective potential for scalar fields on a Schwarzschild background. Note that this expression does not depend on $\xi$. This is because the scalar  curvature is given by $R=8\pi(-\rho+3p)$, and expanding around $r=2M$ we have $\rho\approx 3p$ at leading order (see \eqref{rho1} and \eqref{p1}), so the term $\xi R$ is subleading. On the other hand, near  the throat $f(r_0)\sim O(\sqrt{\hbar}/M)$ so the quantum correction of the effective potential is of order $O(\sqrt{\hbar}/M^3)$ near  the throat, while it is of order $O(\hbar/M^4)$ in general. 

Using this expression for the corrected effective potential, we can \black{now obtain} the quantum corrections at first order in $\hbar$ to the light ring frequencies. 
The computation of these frequencies requires  numerical methods. However, one can obtain a reasonable estimation by using the WKB approximation \cite{iyer-will}.

Let us briefly review the calculation for a Schwarzschild metric. In this framework  the light ring frequencies 
at $0-$th adiabatic order are given by 
\be
\omega_n^2=V(r^*_m)-i\sqrt{-2V''(r^*_m)}(n+\frac12) \, , \label{QNFormula}
\ee
where $r^*_m$ is the value of the tortoise coordinate at which  the potential is maximum, and the primes mean derivatives with respect to $r^*$. \black{In} the case of a Schwarzschild background the maximum of the potential is located at 
\be
r_m=\frac{3 (\lambda -1)+\sqrt{  (9 \lambda +14)\lambda+9}}{2 \lambda }M\, ,
\ee
which for large $l$ \black{tends} to $r_m\sim 3M$. [The case $\lambda=0$ ($l=0$) has to be studied separately, we analyze it at the end of this section]. Using this expression, we can obtain the frequency of the light-ring modes for a scalar perturbation in a classical black hole
\be
\small{\omega_{Sch}^2=\frac{1}{M^2}\left(1-\frac{2}{\tilde r_m}\right)\left(\frac{\lambda}{\tilde r_m^2}+\frac{2}{\tilde r_m^3}\right)-i\left(n+\frac12\right)\frac{2}{M^2\tilde r_m^4} \sqrt{\left(1-\frac{2 }{\tilde r_m}\right) \left(-96 \tilde r_m-10 (3 \lambda -7) \tilde r_m^2+4 (5 \lambda
   -3)  \tilde r_m^3-3 \lambda  \tilde r_m^4\right)}}\, ,
\ee
where $\tilde r_m=r_m/M$. 

\black{Now let us} see how this expression changes if we add quantum corrections at first order in $\hbar$. The corrected effective potential \eqref{corrected_pot} has its maximum at $r=r_m+\frac{\hbar}{M}\epsilon+O(\hbar^2)$, where
\bea 
\epsilon&=&\frac{1}{5040 \pi}  
   \tilde r_m^{-4}  \left(40 +12 (\lambda -1) \tilde r_m-3 \lambda 
   \tilde r_m^2\right)^{-1}\left(1-\frac{2}{\tilde r_m}\right)^{-2}\biggl[-392  (48
   \xi -35)+8  (21 \lambda +3360 \xi -2663)\tilde r_m\nonumber\\
   &-&2  (249 \lambda
   +6384 \xi -5057)\tilde r_m^2 +12  (6 \lambda +168 \xi -59)\tilde r_m^3+3 (89 \lambda -177) 
   \tilde r_m^4+27 (3-5 \lambda )  \tilde r_m^5+18 \lambda  \tilde r_m^6\nonumber\\
   &+&\frac{9}{2}\tilde r_m^5 \left(-32 -9 (\lambda -1) \tilde r_m+2 \lambda 
   \tilde r_m^2\right)\left(1-\frac{2}{\tilde r_m} \right)^2 \log \left(1-\frac{2 }{\tilde r_m}\right)\biggr]\, .
\eea
Using the equation \eqref{QNFormula}  we  obtain the following expression for the corrected frequencies at first order in $\hbar$ 
\bea
\text{Re}[\omega^2]&=&\text{Re}[\omega_{Sch}^2]+\frac{\hbar}{630 \pi M^4}\frac{  336 (\lambda +2) \xi -201 \lambda -560+\tilde r_m \left(-84 (\lambda +3) \xi +13 \lambda
   ^2+42 \lambda +210\right)}{  \lambda  \tilde r_m^8
   \left(1-\frac{2}{\tilde r_m}\right)}+O(\hbar^2)\, .\\
\text{Im}[\omega^2]&=&\text{Im}[\omega_{Sch}^2]-\frac{\hbar}{2520 \pi M^4}\left(n+\frac12\right)  \tilde r_m^{-8} \left(1-\frac{2 }{\tilde r_m}\right)^{-3/2} \left(-96 \tilde r_m-10 (3 \lambda -7) \tilde r_m^2+4 (5 \lambda
   -3)  \tilde r_m^3-3 \lambda  \tilde r_m^4\right)^{-1/2}\nonumber\\
   &\dot&\biggl[1176 (144 \xi -125)+12 
   \tilde r_m (-749 \lambda -25760 \xi +645120 \pi  \tilde r_m \epsilon +23011)\nonumber\\
   &+&4  \tilde r_m^2 (3977 \lambda +52584 \xi +25200 \pi 
   (21 \lambda -121) \tilde r_m \epsilon -45476)
   -2 \tilde r_m^3 (3468 \lambda +31584
   \xi +25200 \pi  (63 \lambda -139) \tilde r_m \epsilon -20771)\nonumber\\
   &+&2  \tilde r_m^4 (-1069
   \lambda +3528 \xi +5040 \pi  (170 \lambda -171) \tilde r_m \epsilon +1956)-9 
   \tilde r_m^5 (-271 \lambda +560 \pi  (77 \lambda -30) \tilde r_m \epsilon +312)\nonumber\\
   &+&9 
   \tilde r_m^6 (-71 \lambda +3360 \pi  \lambda  \tilde r_m \epsilon +30)+54 \lambda 
   \tilde r_m^7\nonumber\\
   &+&\frac{9}{2}\tilde r_m^4  \left(-768-14 (15 \lambda -59) \tilde r_m -6 (47-35 \lambda ) 
   \tilde r_m^2-5 (13 \lambda -6) 
   \tilde r_m^3+6 \lambda  \tilde r_m^4\right) \left(1-\frac{2}{\tilde r_m}\right)\log \left(1-\frac{2}{\tilde r_m}\right)\biggr]+O(\hbar^2)\, .
   \eea

As mentioned above, the \black{case} $l=0$ requires special attention.
   In this case the effective potential has its maximum at
   \be
   r=\frac{8 M}{3}+\frac{\hbar}{430080 \pi  M} (1008 \xi -1767+2048 \log (2))+O(\hbar^2)\, .
   \ee
   Therefore, the corrected frequency at first order in $\hbar$ for $l=0$ is given by
   \be
   \omega^2=\omega_{Sch}^2+\frac{3 \hbar}{286720 \pi  M^2} \left((336 \xi -241) \text{Re}[\omega^2_{Sch} ]-2 i (336 \xi -5) \text{Im}[\omega^2_{Sch} ]\right)
   \ee
    One can see \black{that}, even if the geometry of the spacetime is drastically changed by quantum effects near to the horizon, they do not imply significant corrections to the physical observables in the exterior region.

\subsection{Electromagnetic perturbations}

Let us \black{now study}  the propagation of electromagnetic waves on a general static and spherically symmetric metric given by $ds^2=g_{tt}(r)dt^2 + g_{rr}(r)dr^2+ r^2 d\Omega^2$. The electromagnetic field $F_{ab}$ satisfies the source-free Maxwell equations:
\be
\nabla_aF^{ab}=0, \quad \nabla_a {^*F}^{ab}=0\, ,
\ee
where ${^*F}$ is the Hodge dual of $F$. The second equation is solved with  $ F_{ab}=A_{a, b}-A_{b, a}$, where $A_a$ is the electromagnetic potential, and the problem is reduced to solve the first equation above for the vector field $A_a$. For a spherically symmetric background spacetime we can search for solutions by expanding $A_a$ in the basis of 4-dimensional vector spherical harmonics $(Y_a)_{\ell m}$. Elements of this basis are classified according to their behavior under parity transformations. 
For axial/odd  modes, which have parity $(-1)^{\ell+1}$, the electromagnetic potential can be expanded as
\bea
A^{-}_a(t, r, \theta, \phi)= \sum_{\ell, m} \left[\begin{array}{c}
0 \\
0 \\
\frac{a^{l m}(t, r)}{\sin \theta} \partial_\phi Y_{l m} \\
-a^{l m}(t, r) \sin \theta \partial_\theta Y_{l m}
\end{array}\right]\, ,
\eea
for some (gauge-invariant) coefficients $a^{l m}(t, r)$. Using this ansatz one can check that there is only one nontrivial independent  equation from $\nabla_aF^{ab}=0$. For a static spacetime we can further separate $a_{\ell m}=e^{-i\omega t}\Psi^{-}_{\ell m}(r)$, and the resulting equation can be written as 
\bea
F^2{\Psi^-}''_{\ell m}+FF' {\Psi^-}'_{\ell m}+(\omega^2-V_l){\Psi^-}_{\ell m}=0 \, ,\label{RWforM} 
\eea
where
\bea
&&F(r)=\sqrt{-\frac{g_{tt}}{g_{rr}}}\, ,
\\
&&V_l(r)=-g_{tt}(r)\frac{l(l+1)}{r^2}\, .
\eea
Again, introducing the tortoise coordinate by $\partial_{r^*}=F(r)\partial_r$, one recovers the usual Regge-Wheeler form of the equation, $\partial_{r_*}^2{\Psi^-}_{\ell m} +(\omega^2-V_{\ell}){\Psi^-}_{\ell m}=0$. In particular, for a Schwarzschild background we recover the usual expression.

For polar/even  modes, which have parity $(-1)^{\ell}$, the electromagnetic potential can be expanded as
\bea
A^{+}_a(t, r, \theta, \phi)= \sum_{\ell, m} \left[\begin{array}{c}
f^{l m}(t, r) Y_{l m} \\
h^{l m}(t, r) Y_{l m} \\
k^{l m}(t, r) \partial_\theta Y_{l m} \\
k^{l m}(t, r) \partial_\phi Y_{l m}
\end{array}\right]\, ,
\eea
for some coefficients $f^{l m}$, $h^{l m}$, $k^{l m}$. \black{However, these coefficients are} gauge-dependent.  \black{Let us introduce} the three gauge-invariant combinations $\Psi^+=\sqrt{-g_{tt} g_{rr}}\frac{r^2}{\ell(\ell+1)}(\partial_t h^{l m}- \partial_rf^{l m})$, $\Psi_{1,\ell m}=f^{l m}-\partial_tk^{l m}$ and $\Psi_{2,\ell m}=h^{l m}-\partial_r k^{l m}$ (these combinations are essentially the field components $F_{tr}$, $F_{t\phi}$, $F_{r\phi}$, respectively; the rest of \black{the} field components are redundant)\black{. Using}   Maxwell equations one can conclude, after some work,  that $\Psi^+_{\ell m}$ satisfies the same equation (\ref{RWforM}) as the axial solution $\Psi^-_{\ell m}$, and the rest of \black{the} field variables are determined from it: $\Psi_1=-\frac{\partial_r\Psi^+}{g_{rr}}+\frac{\partial_r(g_{tt}g_{rr})}{2g_{tt}g_{rr}^2}\Psi^+$ and $\Psi_2=\frac{\partial_t\Psi^+}{g_{tt}} $. One can easily check that these results fully solve the system of equations $\nabla_aF^{ab}=0$, and the whole problem reduces to solve (\ref{RWforM}) with suitable boundary conditions. 

The fields $\Psi^{\pm}_{\ell m}$ constitute the two fundamental degrees of freedom per spacetime point of the electromagnetic field. Notice that both fields satisfy exactly the same dynamical equation even when the quantum corrections considered in this paper are included, leading in particular to the usual phenomenon of isospectrality \cite{ReviewVitor}. This could have been guessed in advance from the electric-magnetic duality symmetry of \black{the} source-free Maxwell equations \cite{Adrian}, \black{since} $\Psi^+$ plays the role of the electric field while $\Psi^-$ represents the magnetic degree of freedom.

For the perturbative corrected Schwarzschild metric provided in Sec. III, the first order correction in $\hbar$ to the potential yields

\be \label{corrected_potEM}
V(r)=V_{1}(r)-\frac{\hbar \,l(l+1) }{5040 \pi  M^2 r^7 }\left( \frac{2 M}{f(r)} \left(21 M^2 r^2+40 M^3
   r-14 M^4-36 M r^3+9 r^4\right)+9 r^4 (r-3 M)  \log (f(r))\right)+O\left(\hbar^2\right)\, ,
\ee
where $V_{1}=f(r)\frac{l(l+1)}{r^2}$ is the potential for electromagnetic perturbations on the Schwarzschild metric, and $f(r)=1-\frac{2M}{r}$. As in the scalar case, for $r\to r_0$ the effective potential acquires a nonzero residual value of order $\sqrt{\hbar}=\ell_p$, \black{which is} not present in the classical case. 

Finally, let us analyze the quantum corrections to the light ring frequencies of electromagnetic perturbations, \black{again using} the WKB approximation described above. For the Schwarzschild metric the maximum of the Regge-Wheeler potential is located at $r=3M$, and therefore using the expression \eqref{QNFormula} one obtains
\be
\omega^2_{Sch}=\frac{l(l+1)}{27M^2}-i\frac{2\sqrt{l(l+1)}}{27M^2}(n+\frac12)\, .
\ee
For our (perturbatively) corrected spacetime, at first order in $\hbar$ we obtain the maximum of the potential \eqref{corrected_potEM} at  $r=3M+\frac{\hbar }{90720 \pi  M}(243 \log (3)-20)$. Substituting in \eqref{QNFormula} and expanding in Taylor series we  obtain the following expression for the quantum correction to the light ring frequencies:
\be
\omega^2=\omega^2_{Sch}+\frac{\hbar}{17010\pi M^2}\left(-13\text{Re}[\omega^2_{Sch}]+11 i \,\text{Im}[\omega^2_{Sch}]\right)\, .
\ee
Again, the quantum effects of vacuum polarization do not lead to significant, observable corrections.

\section{Summary and final comments}
The theory of test quantum fields in a given gravitational background is widely regarded as a  useful and fruitful  framework \black{for exploring} quantum fluctuations enhanced by gravity.
This theory can  be further used to analyze the backreaction of these quantum effects  on the spacetime background by looking at the semiclassical Einstein's equations (\ref{Eseq}).
Solving these equations is, however, a very elusive problem and only in very highly symmetric situations one can carry out the computation in a manageable way. A good  example are conformally flat spacetimes with conformal matter fields. In this case $\langle T_{ab} \rangle$ is  essentially  characterized by the conformal anomaly. Another relevant example emerges in two-dimensional dilaton-gravity models coupled to conformal matter.  The conformal anomaly in two dimensions fully determines the quantum stress tensor for a given choice of the vacuum state, thus allowing us  to solve analytically the semiclassical backreaction equations for a relevant class of two-dimensional models \cite{FBI-Fabbri}.  

In this paper we \black{have}  reanalyzed the four-dimensional problem from scratch, focusing on static and spherically-symmetric backgrounds. The general expressions given in \cite{Anderson95}  for the renormalized  stress tensor, when the quantum field lives in static and spherically symmetric spacetimes, \black{represent} a very significant progress,  \black{but} they are still quite involved and unpractical to solve the semiclassical equations. One way to simplify the problem  is to restrict  ourselves to conformal matter and take advantage of the trace anomaly. However, those assumptions (spherical symmetry, staticity and conformal matter) are \black{still} not  sufficient  to reduce the problem to a manageable form, in sharp contrast with the effective two-dimensional case \cite{Fabbri-Navarro, Ho-Matsuo18, Julio}.  To  overcome this difficulty we have introduced an extra simplifying assumption, suggested by well-known  results in the fixed Schwarzschild background.   Since we are mainly interested in the behavior of the geometry in the very near horizon region $r\sim 2M$ (in the macroscopic vicinity of $2M$ one does not expect any significant modification of the classical Schwarzschild geometry)  we have  assumed  the exact relation between $\langle p_t\rangle$ and $\langle p_r\rangle$ in the vicinity of the classical horizon (suggested by the results in the fixed background approach).   
Our findings appear to be essentially insensitive of this assumption.  More precisely, we have numerically checked that the (nonperturbative) backreaction solution obtained with other restrictions (\black{such as} $\langle p_t \rangle = 0$)   
are qualitatively similar to \black{those} described in Sec. (\ref{solution}). Furthermore, our results do not  depend on the particular form of \black{the} conformal matter either (for a massless Dirac field we have obtained results \black{similar to} those for a scalar field). 

One remarkable property of the semiclassical backreaction solution obtained in Secs. \ref{solution} and \ref{extension} is that
the radial function can never reach $0$ (where the classical curvature singularity is located), but rather it has \black{a} minimum \black{on} a timelike surface. This mimics  the throat of an (asymmetric) wormhole, and it  is   located at $r_0 \approx 2M + \mathcal{O} (\sqrt{\hbar})$, where  the red-shift function reaches a very small but nonzero value.\footnote{We note that the power in the dependence on $\hbar$ is different from that obtained in the approach of Ref. \cite{ho-kawai-matsou-yokokura}, for which $r_0 \approx 2M + \mathcal{O} (\hbar)$.  Furthermore, we also have discrepancies in the analytic form of the metric components.} 
Beyond this bouncing surface for the radial function we have found a null curvature singularity at a finite proper-time distance (of order $\mathcal{O}(\sqrt\hbar)$ from the throat).  The overall physical picture qualitatively  agrees with the results  obtained from the purely two-dimensional approach.  This indicates that the two-dimensional approach could be more accurate than it could be expected.

The global picture obtained from this semiclassical framework differs drastically from its counterpart in   classical general relativity, specially regarding the   black hole interior region. Strictly speaking, here the classical horizon disappears and it is replaced by a bouncing timelike surface,  beyond which a null curvature singularity emerges immediately. The underlying reason for this seems to be rooted \black{in} the singular behavior of the renormalized stress tensor at the classical  horizon obtained in the fixed background approach. In light of these results, it looks as if   the original singular behavior of the stress-tensor in the classical horizon manifests itself in the metric in the form of a curvature singularity as a result of the backreaction. We regard this  singularity as a side effect of \black{the assumption of} a pure vacuum solution. 
The presence of matter could tame the singularity if vacuum polarization effects continue to be relevant (as suggested by the results in \cite{Julio-2}) and allow the formation of ECO's.   However in this case the maximum compactness of \black{these} objects is bounded by $2M/r_0 \sim 1- 0.01686\sqrt{\hbar}/(2M)$. This bound is a direct consequence of the fact that the exterior geometry of  ECOs \black{has} to be described by the external portion of our semiclassical solution, and not by the classical Schwarzschild metric.

We have also analyzed potential physical implications of the quantum corrected geometry in the exterior region. In Sec. \ref{perturbations} we have analyzed in detail the scalar and electromagnetic perturbations, paying special attention to the so-called ``light ring frequencies,'' which are the relevant observables in the ringdown  of  binary black holes. 
We have evaluated the corrected light ring frequencies using our predictions for the semiclassical metric, and they differ from their classical counterpart 
by  corrections of order $\mathcal{O}(\hbar/M^2)$. Somewhat not surprising, the drastic modification of the metric around the classical horizon does not lead to observable corrections on these observables, \black{since} these frequencies are  determined by the spacetime curvature around the light-ring. To really probe the quantum corrections around the classical horizon geometry one would need to compute the proper BH quasinormal mode frequencies of the system, which would most likely differ nonperturbatively from the classical BH QNMs. However these observables require \black{the} specification of boundary conditions at the center or surface of the quantum object in question, and this is out of the scope of the present paper.

We plan  to extend this work in several directions. Apart from computing the  QNM frequencies above, our goal is to analyze  the inclusion of  collapsing matter  and the impact of the time-dependent phase on the backreaction effects.  This is indeed a very difficult problem in the four-dimensional arena and requires a separate study.    

Note added: after \black{the} communication of this paper  another work appeared \cite{ABCRG} which confirms some of our conclusions, giving further support to our results.

\section{Acknowledgments}

We would like to thank V. Cardoso for useful discussions as well as comments on the manuscript. We also thank I. García Martínez for helpful collaboration in some numerical calculations.   Financial support was provided by  Spanish Grants No. 
 PID2020-116567GB-C2-1  funded by No. MCIN/AEI/10.13039/501100011033, 
 and No. PROMETEO/2020/079 (Generalitat Valenciana). 
P. B-P. is supported by the Ministerio de Ciencia, Innovaci\'on y Universidades, Ph.D. fellowship, Grant No. FPU17/03712. 
A.D.R. acknowledges support from the NSF grant No. PHY-1806356 and the Eberly Chair funds of Penn State.
We acknowledge  use of some packages of {\bf xAct} for \textit{Mathematica}. 

\appendix

\section{Nature of the singular point $r=r_0$} \label{appendix}

In this appendix will analyze in detail the nature of the singular point $r=r_0$ obtained in Sec. \ref{solution} and will prove that it is a coordinate singularity. We will also see why this singular point does not define a classical horizon.

Let us consider a general metric of the form
\be \label{general_metric}
ds^2=-G(r)dt^2+\frac{dr^2}{F(r)}+r^2d\Omega^2\, ,
\ee
with $G(r)>0$ and $F(r)>0$. Its \black{corresponding} curvature scalar is given by
\be
R=\frac{4 G^2 \left(r F'+F-1\right)+r G  \left(G' \left(r F'+4 F\right)+2 r F 
   G''\right)-r^2 F  \left(G'\right)^2}{2 r^2 G^2}\, .
\ee
In our case $F(r_0)=0$ at the singular point, but $G(r_0)\neq 0$ and their derivatives are not divergent, so the scalar curvature  is finite at this point, and therefore $r=r_0$ is a coordinate singularity. This statement can also be inferred  from a perturbative analysis of the  semiclassical Einstein's equations $R=8\pi\langle T_a^a\rangle$, \black{since} at first order the trace does not diverge ($\langle T_a^a\rangle=\frac{\hbar M^2}{60 \pi^2 r^6}+O(\hbar^2)$).

Another way to confirm this,  and to  assess the impact of the quantum-vacuum polarization on the classical Schwarzschild geometry, is by analyzing the Kretschmann curvature scalar. For a static \black{and} spherically symmetric metric, the explicit expression can be  simplified considerably if we use the TOV equations. It reads
\be
K(r)=16 \left(-\frac{8 \pi  m(r) \langle \rho (r)\rangle }{r^3}+\frac{3 m(r)^2}{r^6}+4 \pi ^2 \left[2 \langle p(r)\rangle \langle \rho (r)\rangle+3 \langle p(r)\rangle ^2+3
  \langle \rho (r)\rangle^2\right]\right) \label{Kscalar}
\ee
Since the renormalized pressure and density are of order $\hbar/f^2$ near the singular point (i.e. numerically of order $\sim 1$ since $f(r_0)\sim\sqrt{\hbar}$), we can see that the Kretschmann scalar does not diverge. In particular, by substituting  the perturbative solution at first order in $\hbar$ into this expression we obtain
\be
K(r)=\frac{48 M^2}{r^6}+\frac{\hbar}{105 \pi
    r^9} \left(\frac{2 M}{r^2f(r)^2} \left(728 M^4-818 M^3 r+212 M^2 r^2+27 M
   r^3-9 r^4\right)-9 r^3 \log \left(1-\frac{2 M}{r}\right)\right)+O\left(\hbar^2\right)\, .
\ee
As mentioned above,  near  the singular point the leading correction to the Kretschmann scalar behaves as $\hbar/f^2$, which tends to $O(1)$ \black{at} this point. Notice that, as compared to the classical Schwarzschild value,  the Kretschmann scalar is expected to receive corrections that are of order $O(\hbar^0)$ in a neighborhood of the singular point, meaning that quantum corrections may be significant for the nearby geometry despite the tiny value of $\hbar$. 

If we  substitute the  equation of state (\ref{eqnstate}) in (\ref{Kscalar}), we see that  \black{the} terms that include the trace anomaly are of order $\hbar$ near the singular point, so for a conformal quantum field  we can further approximate the Kretschmann scalar  as
\be
K(r)\sim16 \left(-\frac{8 \pi  m(r) \langle\rho (r)\rangle }{r^3}+\frac{3 m(r)^2}{r^6}+16 \pi ^2 \langle\rho (r)\rangle^2\right)
\ee

As mentioned above, this coordinate singularity does not define a classical horizon. To check \black{this explicitly}, it is useful to switch to Eddington-Finkelstein coordinates.
Defining the generalized tortoise coordinate as $ dr_*^2=G^{-1}F^{-1}dr^2$ and the advanced time as $v:=t+r_*$, the metric \eqref{general_metric} can be expressed as
\bea
ds^2=-G(r)dv^2+2F^{-1/2}(r)G^{1/2}(r)dv dr + r^2 d\Omega^2\, . \label{EFcoord}
\eea
Notice that $2F^{-1/2}G^{1/2}dv dr=-(-ds^2-Gdv^2+r^2 d\Omega^2)$. Therefore for causal ($ds^2\leq 0$) and future-directed ($dv>0$) curves, $dr<0$ is only possible if $G(r)<0$. If there \black{were} a critical point \black{where} $G(r)=0$, it would define a one-way membrane for radial ($d\Omega=0$) null geodesics, i.e. a horizon. But in our case $G(r)>0$ for all $r\geq r_0$, so there is no horizon in this spacetime.

As a side remark, notice that in sharp contrast to the Schwarzschild metric \black{where} $F(r)=G(r)$, the Eddington-Finkelstein coordinates are not useful to penetrate across the coordinate singularity $r=r_0$, because the metric in these coordinates is not regular. We discuss  the question of how to extend the metric  across $r=r_0$ in Sec. IV.

\end{document}